%% file: tau-b.tex
\documentclass[11pt]{article}

\def\l {\lambda}

\def\r {\rightarrow}

\def\rnot {R\!\!\!/}
\def\bar {\overline}

\def\be {\begin{equation}}
\def\ee {\end{equation}}
\def\bea {\begin{eqnarray}}
\def\eea {\end{eqnarray}}

\def\bc {\begin{center}}
\def\ec {\end{center}}

\begin{document}
\title{
\rightline{\small{hep-ph/0205046}} \vspace*{0.5cm}
\boldmath
\bf Constraints on R-parity violating supersymmetry from 
leptonic and semileptonic $\tau$, $B_d$ and $B_s$ decays\unboldmath}
\author{ {\large\bf Jyoti Prasad Saha}\thanks{Electronic address: 
jyotip@juphys.ernet.in} 
~and~ {\large\bf Anirban Kundu}\thanks{Electronic address: 
akundu@juphys.ernet.in}\\[3mm]
Department of Physics, Jadavpur University, Kolkata 700032, India}

\date{\today}
\maketitle

\begin{abstract}
We put constraints on several products of R-parity violating $\lambda
\lambda'$ and $\lambda'\lambda'$ type couplings from leptonic and 
semileptonic $\tau$, $B_d$ and $B_s$ decays. Most of them are one to two 
orders of magnitude better than the existing bounds, and almost free from 
theoretical uncertainties. A significant improvement of these bounds can 
be made in high luminosity tau-charm or $B$ factories.
\end{abstract}

\noindent PACS number(s): 13.20.He, 13.35.Dx, 12.60.Jv, 11.30.Fs 
\newpage

\centerline{\bf 1. Introduction}

We live in a time when Standard Model (SM) has been vindicated in a number
of experiments and at the same time has left us with a feeling that it is
incomplete, which makes the search for physics beyond SM a holy grail for
most of the particle physics community. We have absolutely no idea what the
manifestation of the new physics will be; this forces us to consider all
sorts of theoretically motivated new physics options. Since one does not
have any experimental signal that definitely points to new physics, the best
one can do is to constrain the parameter space of the new physics models.
These constraints come mostly from experimental data (including astrophysical
ones) but sometimes from theoretical considerations too \footnote{For example,
from the consideration of the stability of the scalar potential.}. 

As long as one does not produce the new particles directly, one has to look
for their indirect effects in low-energy observables. All low-energy data are
more or less consistent with the SM, taking into account the experimental 
and theoretical errors and uncertainties. Thus, one can look at those
observables which may be explained by SM; the trick is to maximize the width
of the window for new physics and put constraints on the parameter space
of such models. Alternatively, one can look at those observables which
are absolutely forbidden or highly suppressed in SM so that one does not
expect any signal; here even one event will signal new physics and in the
absence of any event, the parameter space for new physics may be constrained
from the experimental upper bounds.

Among the new physics options that people consider, supersymmetry (SUSY),
with all its variants, is the most popular one \cite{susy}. 
In the minimal and some 
non-minimal versions of SUSY, the action is so taken as to conserve the
$R$ quantum number defined as $R=(-1)^{3B+L+2S}$, where $B$, $L$ and $S$
stand for baryon number, lepton number and spin of the field respectively. 
This ensures $R=+1$ for all particles and $R=-1$ for all superparticles,
and conservation of $R$ imply that superparticles must occur in pair in
all allowed Feynman vertices. However, R-parity is a discrete symmetry
imposed by hand (to make the parameter space of the model more restricted
and tractable) and one can write a R-parity violating superpotential
of the form
\be
W={1\over 2}\lambda_{ijk}L_iL_jE^c_k+
\lambda'_{ijk}L_iQ_jD^c_k+
{1\over 2}\lambda''_{ijk} U^c_iD^c_jD^c_k
   \label{rpv-superpot}
\ee
which does not violate any gauge symmetry. The factors of $1/2$ take care
of the fact that $\l$ and $\l''$ couplings are antisymmetric in its
first two indices. Such R-parity violating
terms can be motivated from some of the grand unified theories \cite{rpv-gut}.
In eq. (\ref{rpv-superpot})
$L$, $Q$, $U$, $D$ and $E$ denote, respectively, $SU(2)_L$ doublet
lepton and quark superfields, and $SU(2)_L$ singlet up, down and charged
lepton superfields, and $i,j,k$ are generation indices. 
Of course, $B$ and $L$ are
both violated, and to forbid proton decay, one has to keep either $L$-violating
$\lambda$ and $\lambda'$ terms or $B$-violating $\lambda''$ terms, but not
both. There is also a bilinear R-parity violating term of the form 
$\epsilon_iL_iH_2$, which has a lot of interesting phenomenology \cite{bilin},
including possible leptonic flavor violation \cite{abada},
but we will not consider that term in the present paper.

With 45 new couplings (and all of them can theoretically be complex) the
phenomenology is immensely richer, but at the same time less predictive.
There is, however, one major point to be noted: R-conserving SUSY can affect
low-energy observables through loop effects and hence can hardly compete
with SM effects (except in some of the cases where the SM process itself
is loop-induced or, even better, forbidden); R-parity violating (RPV) SUSY,
on the other hand, can show up in tree-level slepton or squark mediated
processes which can successfully compete with the SM. This also ensures that
for comparable coupling strength, RPV SUSY amplitudes are generally orders of
magnitude higher than the R-conserving SUSY amplitudes. 

The individual RPV couplings have been constrained from various low-energy
processes \cite{dreiner}, and upper limits on some of 
the product couplings, including their phases, have also been found
\cite{maalampi, ko, jang, gg-arc, dreiner2, dl}. Often 
one finds that the product coupling is much more constrained than the direct
product of the upper bounds of the individual couplings \footnote{Constraints
coming from $\Delta m_K$, $\Delta m_B$, $\mu\rightarrow 3e$, $\mu\rightarrow
e$ conversion, etc.\ fall in this category.}.  

In this paper we find the upper limits on the products of 
some of the $L$-violating $\lambda$ and $\lambda'$
type couplings coming from rare $\tau$, $B_d$ and $B_s$ decays. 
The leptonic flavour violating processes are forbidden in the SM, 
and the only contributing amplitude comes from RPV SUSY. The expected 
branching ratios (BR) of leptonic flavour conserving $\Delta B=1$ processes 
within SM are 
so much below the experimental numbers (except $B\rightarrow K^{(*)}\ell^+
\ell^-$) that one can safely ignore the SM effects, as well as the R-conserving
SUSY effects, to put bounds on the RPV couplings. Most of these decay modes 
are also theoretically clean and free from any hadronic
uncertainties which plague nonleptonic decays. (The exception is again the
semileptonic process $B\rightarrow K^{(*)}\ell^+ \ell^-$.)

There are three major sources of uncertainty, however. First, the decay 
constants of the neutral mesons, particularly
that of $\eta$, $B_d$ and $B_s$, are yet to be cleanly determined.
Fortunately, the bounds where these decay constants are relevant just
scale with their values; this will be discussed in the appropriate section.
Second, the $B\rightarrow K/\pi$ formfactors allow a slight theoretical
uncertainty. We use the BSW formfactors with oscillator parameter $\omega
=0.5$ GeV \cite{bsw}.
Lastly, the current masses of the light quarks cause the maximum 
uncertainty in $\l\l'$ type couplings; again, the bounds scale with the
quark masses.  The individual bounds in most
of the cases are
fairly weak and the bounds on the product couplings 
that we find are sometimes one to two orders of magnitude better 
than the existing ones. Some of these product couplings have been considered
earlier \cite{ko,jang}; these we update with better experimental numbers.
These updated numbers, too, are vastly improved. We have not discussed
leptonic and semileptonic $D^0$ decays since the bounds
are much weaker than those which one gets from $\mu\rightarrow e$
conversion \cite{maalampi}. This in turn implies that such decay
signals of $D^0$, if observed in present and upcoming colliders, imply
some new physics but not RPV SUSY.

When we say that these bounds are robust, we of course not only mean to
alleviate the theoretical uncertainties of the SM. The fact that leptonic
flavour-changing processes are absolutely forbidden in the
SM imply that these bounds stand no matter what the phases of these RPV
couplings may be. The same is true for all purely leptonic $B$ decays,
but not if there are competing SM
amplitudes (as in $B\r K\ell^+\ell^-$), 
even if one takes the lowest possible SM number and saturates
the experimental data by RPV contribution, which is the standard practice.
The reason is simple: the two amplitudes are coherent and the bounds depend
on the phase of the RPV couplings, while the standard prescription is true
for incoherent amplitudes only. This we show explicitly for the $B\rightarrow
K\ell^+\ell^-$ decays. One should note here that the most conservative
bounds, not necessarily the true ones, come from such incoherent amplitude
summation. 

There is another point that we like to emphasize. Though it is true 
that observation of the SM-forbidden decays would be a definite signal of
new physics, to show convincingly that it is RPV SUSY one needs to find some
correlated signals ({\em e.g.}, enhancement of BRs)
in different decay channels. We briefly discuss
how this can be done for explicit RPV models (spontaneous RPV models have
been discussed in \cite{frank}, including bounds coming from mesonic and 
leptonic flavor violating decays).
If one finds such correlated signals, it will be an almost definite RPV
signal, without the direct observation of superparticles. 
Such signals occur from the fact that same four-Fermi operators lead to 
different final states. Among leptonic and semileptonic $B$ decays, one
may mention the correlated channels (i) $B_d\r e\mu$ and $B_d\r \pi e\mu$;
(ii) $B_s\r \ell_i^+\ell_j^-$ and $B_d\r K\ell_i^+\ell_j^-$. More correlated
channels are to be found in nonleptonic B decays, which will be discussed in
a future paper \cite{gg-ad-ak}.  
In the absence of any correlated decay channel, the next best thing is to
observe the decay distribution of the final state particles, since RPV SUSY
has a different Lorentz structure from that of the SM. One may, for
example, study the angular distribution of final state leptons in $B\r K\ell^+
\ell^-$ decays --- the crucial fact is that tree-level RPV has a Lorentz
structure of the form $(V-A)\otimes (V+A)$. We do not go into any detailed
discussion of this issue in the present work.

We consider only nonzero L-violating $\lambda$ and
$\lambda'$ type couplings. Though it is true that leptonic
$\tau$ decays can be mediated by $\lambda$ type couplings alone, the
individual bounds on these couplings are much tighter than one may hope to
get from such decays. Thus, $\tau\rightarrow 3$ lepton (there can
be six different combinations) processes give a fairly weak bound ($\sim
{\cal O}(1)$) on the relevant $\l$-type couplings. We do not consider such
processes further; if they are observed in near future, their explanation
must lie somewhere else. 
It is obvious that $\lambda''$ type couplings cannot mediate leptonic
and semileptonic decays.

To construct four-Fermi operators from $\lambda$ and $\lambda'$ type
couplings that mediate such
semileptonic and leptonic $\tau$ and $B$ decays,
one needs to integrate out the squark or the slepton propagator.
Both the couplings coming 
in the product may be complex; one is free to absorb the phase of one coupling
in the sfermion propagator but the other remains, making the overall coupling
responsible for the process a complex one in general. However, since all
these processes are one-amplitude (there is no SM counterpart) no scope of
CP-violation exists; one only observes the nonzero branching ratios. By
the same argument, we can take all couplings to be real without any loss
of generality. The only exception to this statement, {\em viz.}
$B\r K(^*)\ell^+\ell^-$, will be dealt in proper place. 

The paper is arranged as follows. In the next section, we discuss the
formalism for, and the bounds coming from, semileptonic $\tau$ decays.
Section 3 discusses leptonic $B_d$ and $B_s$ decays, while section 4 is
on semileptonic decays of these mesons. We summarize and conclude in
section 5. 

\vspace*{1cm}

\centerline{\bf 2. \boldmath $\tau$ \unboldmath decays}

All the processes that we consider involve terms in the RPV Hamiltonian
with two leptons and two quarks as external fields. The Hamiltonian can
be written as
\bea
{\cal H}_{\rnot} &=& -A_{jklm}(\bar\ell_j(1+\gamma_5)\ell_k)(\bar{d_m}
(1-\gamma_5)d_l) + {1\over 2}B_{jklm}(\bar{\ell_j}\gamma^\mu(1-\gamma_5)
\ell_l)(\bar{d_m}\gamma_\mu(1+\gamma_5)d_k)\nonumber\\
&{ }& -{1\over 2}C_{jklm}(\bar{\ell_j}\gamma^\mu(1-\gamma_5)\ell_l)
(\bar{u_k}\gamma_\mu(1+\gamma_5)u_m) + h.c.
\eea
where 
\be
A_{jklm} = \sum_{i=1}^3{\l^*_{ijk}\l'_{ilm}\over 4m_{{\tilde\nu_i}^2} };\ \ 
B_{jklm} = \sum_{i=1}^3
        {{\l'}^*_{jik}\l'_{lim}\over 4m_{{\tilde{u_L}_i}^2} };\ \ 
C_{jklm} = \sum_{i=1}^3
        {{\l'}^*_{jki}\l'_{lmi}\over 4m_{{\tilde{d_R}_i}^2} }. 
\ee
To put bounds, we will assume only one of the $A$, $B$ or $C$ terms to
be nonzero so that there is no interference effect between different RPV
couplings. We will also take only one sfermion generation index $i$ to
be nonzero at a time. 

Note that if the final state consists of two down-type quarks, both $A$
and $B$ terms may contribute, whereas for two up-type quarks, only the
$C$ term comes. For mesons like $\pi^0$, $\rho^0$ or $\eta$ in the final
state, all the three terms may be important. 

One thing that we do not consider is the running of the RPV couplings
between the sfermion scale and the low-energy scale. The corrections are
electroweak in origin and can be safely neglected. The only QCD correction
may occur between the two quark fields; for $\tau$ decays, this pair
hadronizes, absorbing all such uncertainties in the decay constant. The same
is true for leptonic $B$ or $D$ decays. For semileptonic meson decays
the formfactors are supposed to take care of these short-distance
corrections. However, this effect is important when we have four
quarks as external fields and may contribute a multiplicative factor of
$\sim 2$ to the effective Hamiltonian \cite{singer}.

The generic process $\tau\rightarrow\ell + M$ is lepton-flavour violating
and does not occur in the SM. Strong experimental upper limits 
exist on at least fourteen modes that we consider here: $\ell=e,\mu$
and $M=\pi,\rho,\eta,K^0,{K^0}^*,\overline{{K^0}^*},\phi$. Such modes are
fairly clean from a theoretical point of view. In RPV SUSY, all such
processes can occur with squark or sneutrino propagator mediating
the decay. Since the squark is very heavy, we do not consider any QCD
effect that may take place between the squark and the final state quarks.

The only uncertainty, albeit small, appears in the decay constants of
the neutral mesons.  Our values for the decay constants are (in GeV) \cite{ff}:
\be
f_\pi = 0.132,\ f_\rho=0.216,\ f_K=0.161,\ f_{K^*}=0.214,\ f_\phi=0.237.
\ee
The decay constants for $\eta$ (and $\eta'$) are obtained from the decay
constants of the octet and singlet mesons $f_8=1.34f_\pi$ and $f_1=1.10
f_\pi$ by a rotation:
\bea
f_\eta^u&=&f_\eta^d=f_8\cos\theta/\sqrt{6}-f_1\sin\theta/\sqrt{3},\nonumber\\ 
f_\eta^s&=&-2 f_8\cos\theta/\sqrt{6}-f_1\sin\theta/\sqrt{3},\nonumber\\ 
f_{\eta'}^u&=&f_8\sin\theta/\sqrt{6}+f_1\cos\theta/\sqrt{3},\nonumber\\ 
f_{\eta'}^s&=&-2 f_8\sin\theta/\sqrt{6}+f_1\cos\theta/\sqrt{3},
\eea
where the angle $\theta$ is estimated to be about $-22^\circ$. 
The current quark masses are taken to be $m_d=10$ MeV and $m_s=200$
MeV.

Following the standard practice, we assume only one product of RPV coupling 
to be nonzero at a time. This eliminates the need to consider their phases
and signs; without any loss of generality, we can assume all products to be
real and positive since only the square of the absolute magnitude of the
product enters in the expression for the decay width. If the final state
quarks are of charge $-1/3$, the mediating squark must be a `left-handed'
up-type one, and if the quarks of charge $+2/3$, the mediating squark
is a `right-handed' down-type one. Note that if two or more products are
simultaneously nonzero, there can in principle be an interference effect
for $M=\pi,\rho$ and $\eta$, and it becomes imperative to consider the
signs and the phases of the product couplings. We neglect this complexity. 

The experimental 90\% CL upper limits on the BRs of
various $\tau$ decay modes are taken from \cite{pdg2000}:
\bea
Br(\tau\r e\pi) < 3.7\times 10^{-6}\ \ \ 
&{ }& Br(\tau\r\mu\pi) < 4.0\times 10^{-6}\nonumber\\
Br(\tau\r e\eta) < 8.2\times 10^{-6}\ \ \ 
&{ }& Br(\tau\r\mu\eta) < 9.6\times 10^{-6}\nonumber\\
Br(\tau\r eK^0) < 1.3\times 10^{-3}\ \ \ 
&{ }& Br(\tau\r\mu K^0) < 1.0\times 10^{-3}\nonumber\\
Br(\tau\r e\rho) < 2.0\times 10^{-6}
\ \ \ &{ }& Br(\tau\r\mu\rho) < 6.3\times 10^{-6}\nonumber\\
Br(\tau\r e\phi) < 6.9\times 10^{-6}
\ \ \ &{ }& Br(\tau\r\mu\phi) < 7.0\times 10^{-6}\nonumber\\
Br(\tau\r e{K^*}^0) < 5.1\times 10^{-6}
\ \ \ &{ }& Br(\tau\r\mu{K^*}^0) < 7.5\times 10^{-6}\nonumber\\
Br(\tau\r e\overline{{K^*}^0}) < 7.4\times 10^{-6}
\ \ \ &{ }& Br(\tau\r\mu\overline{{K^*}^0}) < 7.5\times 10^{-6}.
\eea

The nonzero $\l\l'$ type couplings can mediate only $\tau\r\ell + P$ type
decays where $P$ is a generic pseudoscalar meson. That the production of
vector mesons is forbidden is evident from the Lorentz structure of the
corresponding four-fermi hamiltonian. The decay width can be written as
\be
\Gamma(\tau\r \ell_i+P[\equiv \bar{q_j} q_k]) = 
{(m_\tau^2+m_{\ell_i}^2-m_P^2) C(m_\tau,m_{\ell_i},m_P) F_P\over
128\pi m_\tau^3 {\tilde m}^4 }|\l_{ni3}\l'_{njk}|^2
\ee
where
\be
C(m_1,m_2,m_3) = \sqrt{m_1^4+m_2^4+m_3^4-2m_1^2m_2^2-2m_1^2m_3^2
-2m_2^2m_3^2},
   \label{callan}
\ee 
\be
F_P = {m_P^4 f_P^2\over (m_{q_j}+m_{q_k})^2},
\ee
and $\tilde m$ denote the slepton mass. Note that due to our definition of
the decay constants, the above expression is to be multiplied by a factor 
of 1/2 only if there is a $\pi^0$ in the final state. The combination
$\l_{ni3}\l'_{njk}$ can be replaced by $\l_{n3i}\l'_{nkj}$ which will 
generate the same decay and hence the same bound applies to both these
combinations.

\input{tab.1}

The bounds are listed in table 1. Note that most of the numbers are at the
same order of magnitude as the previous bounds coming from direct product 
of the individual couplings, though some have been improved\footnote{
Previous bounds indicate the numbers coming from direct product of individual
bounds, or, in certain cases, bounds on the product coming from a different
process. The bounds which are updated have not been considered as previous
bounds.}. We show 
only the best numbers; for example, the decay $\tau\r e+\pi^0$ puts a 
weaker bound on $\l_{231}\l'_{211}$ than that coming from $\tau\r
e+\eta$, and hence is not shown separately. Here, and in all other cases, we
take all squarks and sleptons to be degenerate at 100 GeV, and the bounds
scale in a simple way: $(m_{\tilde q,\tilde\ell}/100~{\rm GeV})^2$.
For squarks of the first two generation this number is not allowed
but used just as a benchmark value; for the lighter stop 100 GeV is
still allowed, and light sbottom is of current phenomenological interest. 
Thus, as far as squark-mediated processes are concerned, 
the more realistic bounds for the first two
generation of squarks should be the number quoted in our tables multiplied
by a factor of $\sim 10$. 

Though we have relied on published experimental numbers only to obtain
the bounds, let us also quote the Belle numbers for the modes $\tau\r eK^0$
and $\tau\r\mu K^0$ \cite{belle-0120}:
\be
Br(\tau\r eK^0) < 1.8\times 10^{-6},\ \ \
Br(\tau\r\mu K^0) < 1.8\times 10^{-6}.
    \label{belle-tau}
\ee
If we use the Belle data, all the eight $\l\l'$ type
products in table 1 where the final state is
either $eK^0$ or $\mu K^0$ have upper bounds of $1.7\times 10^{-3}$, which is
better or compatible to the previous numbers.

The generic couplings $B$ and $C$ (eq.\ \ref{rpv-superpot}) mediate the
decay $\tau\r\ell + M$ where $M$ can be either pseudoscalar or vector.
The expression for the decay widths in these two cases are as follows:
\bea
\Gamma(\tau\r \ell_i+P[\equiv \bar{q_j} q_k]) &=& 
{f_P^2 C(m_\tau,m_{\ell_i},m_P) P_0(m_\tau,m_{\ell_i},m_P)\over
512\pi{\tilde m}^4m_\tau^3} |\l'_{3nk}\l'_{inj}|^2\nonumber\\
\Gamma(\tau\r \ell_i+V[\equiv \bar{q_j} q_k]) &=& 
{f_V^2 C(m_\tau,m_{\ell_i},m_V) V_0(m_\tau,m_{\ell_i},m_V)\over
512\pi{\tilde m}^4m_\tau^3} |\l'_{3nk}\l'_{inj}|^2
\eea
where $C(m_\tau,m_{\ell_i},m_V)$ is defined in eq.\ (\ref{callan}) and 
\bea
P_0(x,y,z)&=& (x^2-y^2)^2 - z^2 (x^2+y^2)\nonumber\\
V_0(x,y,z)&=& z^2(x^2+y^2-z^2) + (x^2-y^2)^2 - z^4
\eea
The combination $\l'_{3nk}\l'_{inj}$ appears if both $q_j$ and $q_k$ are
down-type quarks. In this case the mediating squark is $\tilde {u_n}_L$.
If the final-state quarks are up-type, the combination that appears is
$\l'_{3jn}\l'_{ikn}$, with rest of the formula remaining unchanged.
Again, the expressions are to be multiplied by a factor of 1/2 if one has
$\pi^0$ or $\rho^0$ in the final state.

We list only the best bounds coming from these processes in table 2. Note
that the best bounds always come from those decays where a vector meson
is involved (if we do not consider the Belle data in eq.\ \ref{belle-tau})
in the final state; this is solely due to the better limits
on the decay modes. Since some of the $\l'$ type couplings have weak
individual bounds, most of our bounds on the product couplings 
are one to two orders of magnitude improvement over the previous bounds. 

Consideration of Belle data in eq.\ (\ref{belle-tau}) shows that the couplings
$\l'_{1i2}\l'_{3i1}$ and $\l'_{2i2}\l'_{3i1}$ both have upper bounds 
of $2.3\times 10^{-3}$ coming from $\tau\r eK^0$ and $\tau\r \mu K^0$ 
respectively, all other bounds remaining unchanged. 

\input{tab.2}

Let us, at this point, highlight certain features of the analysis:
\begin{itemize}
\item Mere observation of a single event in any of the decay modes will
signal new physics.
\item However, in a dedicated tau-charm factory, one may hope to observe
more events in different channels if the RPV couplings are close to their
present bounds obtained in tables 1 and 2.
\item With only one nonzero $\l'\l'$ product, one should observe signals
in different modes corresponding to the same quark-level subprocess: {\em
e.g.}, $\tau\r e\pi, e\eta, e\rho$ all should show some anomalous
behaviour. This is another example of correlated channels and is
important to establish the nature of the new physics.
If no such signals are observed in the $\tau\r \mu + M$ channels,
flavour-specific nature of the new physics will all the more be established.
\item If signals are observed in the pseudoscalar channels but not in the
vector channels, that will probably indicate the presence of a $\l\l'$
type coupling. If the opposite happens, RPV explanation of new physics
will be more difficult to sustain; probably one has to invoke cancellation
between different RPV contributions.
\item Finally, note that though the individual $\l$ or $\l'$ type couplings
are $L$-violating, the overall four-fermi hamiltonian conserves $L$ (it
violates leptonic flavour). Thus, one cannot explain processes like
$\tau^-\r\ell^+M_1^-M_2^-$ with RPV.
\end{itemize}

\vspace*{1cm}

\centerline{\bf 3. Leptonic \boldmath $B_d$ and $B_s$ \unboldmath decays}

The leptonic flavour-violating decays $B_{d,s}\r \ell_i^\pm\ell_j^\mp$ ($i
\not= j$) are forbidden in the SM, and flavour-conserving decays ($i=j$)
are so suppressed (except for $\ell=\tau$ which we do not consider anyway)
that we can take them to be almost forbidden to a very good extent. Thus,
the entire amplitude, if nonzero, is solely due to new physics. In RPV 
models, both slepton-mediated $\l\l'$ type and squark-mediated $\l'\l'$
type interactions can cause such purely leptonic decays. As already
stressed, the bounds are robust in the sense that they are free from any
theoretical uncertainties (except for the decay constants of $B_d$ and
$B_s$), and do not depend on the phase of the RPV couplings.

The decay width of $B_{d,s}\r \ell_l^-\ell_m^+$ is given by
\be
\Gamma(B_{q_i}\r \ell_l^-\ell_m^+) = {f_{B_{q_i}}^2 \over 16\pi {\tilde m}^4
M_{B_{q_i}}^3 } C(M_{B_{q_i}},m_{\ell_l},m_{\ell_m}) P_1(M_{B_{q_i}},
m_{\ell_l},m_{\ell_m}) |\l_{nlm}\l'_{ni3}|^2 
           \label{llp-b}
\ee
or
\be
\Gamma(B_{q_i}\r \ell_l^+\ell_m^-) = {f_{B_{q_i}}^2 \over 256\pi {\tilde m}^4
M_{B_{q_i}}^3 } C(M_{B_{q_i}},m_{\ell_l},m_{\ell_m}) P_2(M_{B_{q_i}},
m_{\ell_l},m_{\ell_m}) |\l'_{lni}\l'_{mn3}|^2. 
           \label{lplp-b}
\ee
Here $P_1$ and $P_2$ are given by
\bea
P_1(x,y,z) &=& x^4 - x^2y^2 - x^2 z^2,\nonumber\\
P_2(x,y,z) &=& x^2 (y^2+z^2) - (y^2-z^2)^2
\eea
and the $C$-function is defined in eq.\ (\ref{callan}). The generic slepton
or squark mass is denoted by $\tilde m$. 
Note that the expression $\l_{nlm}\l'_{ni3}$ can be replaced by 
$\l_{nml}\l'_{n3i}$ in eq.\ (\ref{llp-b}) and $\l'_{lni}\l'_{mn3}$
can be replaced by $\l'_{ln3}\l'_{mni}$ in eq.\ (\ref{lplp-b}).
Since we consider only one combination to be present at a time, the same
bound applies to all such combinations. 

\input tab.3

We present our numbers for $\l\l'$ couplings in table 3 and $\l'\l'$
couplings in table 4. Our input parameters are the experimental
90\% CL upper bounds on the BRs of the following modes 
\cite{pdg2000,cleo-bergfeld}
\bea
Br(B_d\r e^+e^-) < 8.3\times 10^{-7};\ \ \ 
&{ }& Br(B_d\r \mu^+\mu^-) < 6.1\times 10^{-7};\nonumber\\
Br(B_d\r e^\pm\mu^\mp) < 1.5\times 10^{-6}; \ \ \ 
&{ }&Br(B_d\r e^\pm\tau^\mp) < 5.3\times 10^{-4};\nonumber\\
Br(B_d\r \mu^\pm\tau^\mp) < 8.3\times 10^{-4};\ \ \ 
&{ }&Br(B_s\r e^+e^-) < 5.4\times 10^{-5};\nonumber\\
Br(B_s\r \mu^+\mu^-) < 2.0\times 10^{-6};\ \ \ 
&{ }&Br(B_s\r e^\pm\mu^\mp) < 6.1\times 10^{-6}.
\eea
Furthermore, we take the decay constants of both $B_d$ and $B_s$ to be
200 MeV; the bounds just scale as $(f_{B_{d,s}}/200$ MeV$)^2$. 

We also have Belle numbers for some of these modes \cite{belle-0127}:
\be
Br(B_d\r e^+e^-) < 6.3\times 10^{-7};\ \ 
 Br(B_d\r \mu^+\mu^-) < 2.8\times 10^{-7};\ \ 
Br(B_d\r e^\pm\mu^\mp) < 9.4\times 10^{-7}.
   \label{belle-blep}
\ee
If one takes these bounds, which are yet to be published, into account, 
some of the numbers in tables 3 and 4 get modified. In table 3, the $\l\l'$
combinations coming from $B_d\r e^+e^-$ have upper bounds of $1.5\times 
10^{-5}$ instead of $1.7\times 10^{-5}$; those coming from $B_d\r\mu^+\mu^-$
and $B_d\r \mu^\pm e^\mp$
have upper bounds of $1.0\times 10^{-5}$ and $1.8\times 10^{-5}$ respectively. 
In table 4, the bounds on the $\l'\l'$ combinations coming from 
$B_d\r\mu^+\mu^-$ and $B_d\r \mu^\pm e^\mp$ are modified to $1.4\times
10^{-3}$ and $3.7\times 10^{-3}$ respectively. Obviously, these bounds
scale as the square root of the upper bounds on respective branching
fractions. 

\input tab.4

The present $e^+e^-$ B factories should improve these bounds by one
order of magnitude at the end of their run, if such modes are not
observed. The hadronic machines or high-luminosity $e^+e^-$ machines
like the projected SuperBaBar \cite{hitlin} should see a large number
of such leptonic decays if the actual values of the couplings are
anywhere near the present bounds. Since the same couplings cause
both $B_d\r \ell_i\ell_j$ and $B\r \pi\ell_i\ell_j$ (the same is true
for $B_s\r ell_i\ell_j$ and $B\r K\ell_i\ell_j$) decays, a simultaneous
signal is expected.

\vspace*{1cm}

\centerline{\bf 4. Semileptonic decays: 
\boldmath $B\r Ke^+e^-, B\r K\mu^+\mu^-, B\r Ke^\pm\mu^\mp,
B\r\pi e^\pm\mu^\mp$\unboldmath}

All three collaborations CLEO, BaBar and Belle have set upper
limits on the BRs of the abovementioned semileptonic modes (and also
modes with a vector meson in the final state) \cite{cleo,babar,belle,cleo2}. 
% \bibitem{cleo} hep-ex/0106060. 
% \bibitem{babar} hep-ex/0107026. 
% \bibitem{belle} K. Abe {\em et al}, Phys. Rev. Lett. 88, 021801 (2002). 
% \bibitem{cleo2} K.W. Edwards {\em et al}, hep-ex/0204017. 
In fact, the modes
$B\r K\mu^+\mu^-$ (and $B\r Ke^+e^-$ at less than $3\sigma$) 
have been observed by Belle \cite{belle}. The present
status is as follows:
\bea
Br (B\r K\ell^+\ell^-) &<& 0.6\times 10^{-6} ~(BaBar), \ \ 1.49\times
10^{-6}~(CLEO) \ \ 
(\ell=e/\mu)\nonumber\\ 
Br (B\r Ke^+e^-) &=&(0.48^{+0.32+0.09}_{-0.24-0.11})\times 10^{-6}~(Belle)
\nonumber\\ 
Br (B\r K\mu^+\mu^-)&=& 
(0.99^{+0.40+0.13}_{-0.32-0.14})\times 10^{-6}~(Belle)\nonumber\\  
Br (B\r he^\pm\mu^\mp) &<& 1.6\times 10^{-6}~(CLEO), \ \ (h=K/\pi).
\eea

These numbers are at the same ballpark as the SM expectations. Unfortunately,
the SM expectations are not precise; at least three different groups
quoted three different numbers, which are all mutually compatible, but
differ in their upper and lower limits \cite{ali,greub,melikhov}. 
For example, the predicted BR for the mode $B\r K\mu^+\mu^-$ is
(i) $(0.57^{+0.16}_{-0.10})\times 10^{-6}$ \cite{ali};
(ii) $(0.33\pm 0.07)\times 10^{-6}$ \cite{greub};
(iii) $(0.42\pm 0.09)\times 10^{-6}$ \cite{melikhov}.
As we
know, the bounds on new physics depend sensitively on the theoretical
predictions. Moreover, the situation is complicated since we should not
neglect the SM amplitudes (apart from the lepton flavour-violating cases).

We take a compromising approach: in table 5, we quote the bounds on the
relevant RPV couplings assuming (i) that the SM expectation
is at its lowest possible value predicted by \cite{ali}; (ii) that the
experimental number is at its highest possible value; (iii) that the
difference is saturated by RPV contribution; (iv) and, most important of
all, that the total SM amplitude and the RPV amplitude add {\em
incoherently}. This is certainly not true; for that, in table 6, we show
how the bounds relevant for a particular decay mode ({\em viz.},
$B\r K\mu^+\mu^-$) change if (i) the RPV amplitude adds constructively
to the SM one; (ii) the RPV amplitude adds destructively with the SM one
(for this, we take the highest possible SM prediction and the lowest
possible experimental number); and (iii) if we use the theoretical numbers
in \cite{greub} or \cite{melikhov}. We draw the attention of the reader
to the huge fluctuation of the RPV bounds. It is, however, heartening to note
that the most conservative bounds come from incoherent amplitude summation,
so that the bounds quoted in table 5 are robust as far as the phases in
the RPV couplings are concerned. This analysis also shows that to get
any meaningful signal of new physics from these modes, which is certainly
feasible in not-too-distant future, one must minimize the theoretical
as well as the experimental uncertainties. Another signal of new physics,
which we do not investigate here, is the forward-backward asymmetry of
the final state leptons.

\input tab.5

For our analysis, we take the most stringent experimental numbers, {\em
viz.}, the numbers quoted by BaBar for $B\r K\ell^+\ell^-$ ($\ell=e,\mu$)
and the numbers quoted by CLEO for $B\r Ke\mu, \pi e\mu$. The SM prediction
for the first two modes can be as low as $0.47\times 10^{-6}$ \cite{ali}
and zero for the last two; with incoherent addition of SM and RPV amplitudes,
one gets the bounds on $\l'\l'$ type products as shown in table 5. We do
not show the $\l\l'$ bounds as they are at least one order of magnitude weaker
than those obtained from leptonic $B_d$ and $B_s$ decays, apart from
being much less robust. We have also used the BSW formfactors \cite{bsw}
for $B\r K$ and $B\r\pi$ transitions and taken $F_0=F_1$ with minimal loss
of accuracy: $F_0^{B\r\pi}(0)=0.39$ GeV, $F_0^{B\r K}(0)=0.42$ GeV.

Though the bounds in table 5 appear to be better than those in table 4,
we warn the reader that they depend sensitively to the experimental
number as well as the scheme for theoretical prediction. 
This is evident from
table 6. Thus, any future analysis with these product couplings should
use the numbers quoted in table 4.

Is it possible to find other channels where one finds such RPV signals?
If only one product coupling is nonzero, there are very few product
couplings constrained here that may contribute to nonleptonic $B$ decays.
Such couplings are: (i) $\l'_{211}\l'_{213}$, mediating $b\r u\bar u d$
and $b\r d\bar d d$, and hence $B\r \pi\pi$ (this coupling also affects
the $B_d$-$\bar{B_d}$ box diagram and thus the mixing-induced CP
asymmetry coming from $B_d$ decays); (ii) $\l'_{212}\l'_{213}$ and
$\l'_{112}\l'_{113}$, mediating
$b\r u\bar u s$ and $b\r d\bar d s$; (iii) $\l'_{221}\l'_{223}$, mediating
$b\r c\bar c d$ and $b\r s\bar s d$; (iv) $\l'_{122}\l'_{123}$ and
$\l'_{222}\l'_{223}$, giving rise to $b\r c\bar c s$ and $b\r s\bar s s$
(and hence $B_d\r J/\psi K_S$ and $B_d\r \phi K_S$). However, these
processes are slepton mediated whereas the leptonic and semileptonic $B$
decays discussed here are squark mediated, and hence the numbers quoted
here should be scaled by $(m_{\tilde q}/m_{\tilde\ell})^2$. The
conclusion is that one should look for any unusual change in CP
asymmetry and/or branching fractions in the abovementioned channels
to get a supporting evidence for RPV SUSY (though, their absence does not
rule out the model).

\input tab.6

\vspace*{1cm}

\centerline{\bf 5. Summary and Conclusions}

In this paper we found the bounds on $\l\l'$ and $\l'\l'$ type
product couplings coming from leptonic and semileptonic $\tau$ and
$B$ decays. Most of our bounds are robust and one to two orders of 
magnitude better than the existing bounds. Some of the updated bounds
also have major improvement. 

The bounds are only on the magnitude of the product couplings. The phase
is irrelevant apart from the semileptonic penguin decays of $B_d$ mesons.
For the latter the most conservative bounds come from incoherent
amplitude summation but depend sensitively on the theoretical prediction
for SM BRs. 

After this work was communicated a paper came to the archive where some of
these bounds have been discussed \cite{akeroyd}.

\vspace*{1cm}

\centerline{\bf{Acknowledgements}}

The work of AK was supported by BRNS, Govt.\ of India (Project No.\ 
2000/37/10/BRNS) and UGC, Govt.\ of India (Project No.\ F.10-14/2001
(SR-I)). JPS thanks CSIR, India, for a junior research fellowship.

\end{document}

%% file: tab.1
%\documentstyle[11pt]{article}
%\pagestyle{empty}
%\oddsidemargin=-0.5 true in
%\begin{document}
\def\ksb {\overline{{K^0}^*}}
\def\kst {{K^0}^*}

\def\r{\rightarrow}
\begin{table}[htbp]
\begin{center}
\begin{tabular}{||c|c|c|c||c|c|c|c||}
\hline
$\lambda\lambda'$ & Final & Bound & Previous&
$\lambda\lambda'$ & Final & Bound & Previous\\
 & state & & bound &
 & state & & bound\\
\hline
(123)(111)& $\mu\eta$& $5.3\times 10^{-4}$ & $2.5\times 10^{-5}$&
(123)(121)& $\mu K^0$& $4.1\times 10^{-2}$ & $1.0\times 10^{-3}$\\
(123)(122)& $\mu\eta$& $1.0\times 10^{-2}$ & $2.1\times 10^{-3}$&
(123)(211)& $e\eta$& $4.9\times 10^{-4}$ & $2.9\times 10^{-3}$\\
(123)(221)& $e K^0$& $4.7\times 10^{-2}$ & $8.8\times 10^{-3}$&
(123)(222)& $e\eta$& $9.3\times 10^{-3}$ & $1.0\times 10^{-2}$\\
(131)(111)& $e\eta$& $4.9\times 10^{-4}$ & $3.2\times 10^{-5}$&
(131)(112)& $e K^0$& $4.7\times 10^{-2}$ & $1.3\times 10^{-3}$\\
(131)(122)& $e\eta$& $9.3\times 10^{-3}$ & $2.7\times 10^{-3}$& 
(132)(111)& $\mu\eta$& $5.3\times 10^{-4}$ & $3.2\times 10^{-5}$\\
(132)(112)& $\mu K^0$& $4.1\times 10^{-2}$ & $1.3\times 10^{-3}$&
(132)(122)& $\mu\eta$& $1.0\times 10^{-2}$ & $2.7\times 10^{-3}$\\
(133)(311)& $e\eta$& $4.9\times 10^{-4}$ & $6.6\times 10^{-4}$& 
(133)(321)& $e K^0$& $4.7\times 10^{-2}$ & $3.1\times 10^{-3}$\\
(133)(322)& $e\eta$& $9.3\times 10^{-3}$ & $1.3\times 10^{-3}$& 
(231)(211)& $e\eta$& $4.9\times 10^{-4}$ & $4.1\times 10^{-3}$\\
(231)(212)& $e K^0$& $4.7\times 10^{-2}$ & $4.1\times 10^{-3}$& 
(231)(222)& $e\eta$& $9.3\times 10^{-3}$ & $1.5\times 10^{-4}$\\
(232)(211)& $\mu\eta$& $5.3\times 10^{-4}$ & $4.1\times 10^{-3}$& 
(232)(212)& $\mu K^0$& $4.1\times 10^{-2}$ & $4.1\times 10^{-3}$\\
(232)(222)& $\mu\eta$& $1.0\times 10^{-2}$ & $1.5\times 10^{-2}$& 
(233)(311)& $\mu\eta$& $5.3\times 10^{-4}$ & $7.7\times 10^{-3}$\\
(233)(321)& $\mu K^0$& $4.1\times 10^{-2}$ & $3.6\times 10^{-2}$& 
(233)(322)& $\mu\eta$& $1.0\times 10^{-2}$ & $3.6\times 10^{-4}$\\
\hline
\end{tabular}
\caption{Bounds on $\lambda\lambda'$ type products from $\tau\r \ell+M$
decays.}
\end{center}
\end{table}
%\end{document}
%

%% file: tab.2
%\documentstyle[11pt]{article}
%\pagestyle{empty}
%\oddsidemargin=-0.5 true in
%\begin{document}
\def\ksb {\overline{{K^0}^*}}
\def\kst {{K^0}^*}
\begin{table}[htbp]
\begin{center}
\begin{tabular}{||c|c|c|c||c|c|c|c||}
\hline
$\lambda'\lambda'$ & Final & Bound & Previous&
$\lambda'\lambda'$ & Final & Bound & Previous\\
 & state & & bound &
 & state & & bound \\
\hline
(111)(311)& $e\rho$& $2.5\times 10^{-3}$ & $5.7\times 10^{-5}$&
(111)(312)& $e\ksb$  & $3.5\times 10^{-3}$ & $5.7\times 10^{-5}$\\
(112)(311)& $e\kst$& $2.9\times 10^{-3}$ & $2.3\times 10^{-3}$&
(112)(312)& $e\rho$& $2.5\times 10^{-3}$ & $2.3\times 10^{-3}$\\
(113)(313)& $e\rho$& $2.5\times 10^{-3}$ & $2.3\times 10^{-3}$&
(121)(321)& $e\rho$& $2.5\times 10^{-3}$ & $22.4\times 10^{-3}$\\
(121)(322)& $e\ksb$  & $3.5\times 10^{-3}$ & $22.4\times 10^{-3}$& 
(122)(321)& $e\kst$& $2.9\times 10^{-3}$ & $22.4\times 10^{-3}$\\
(122)(322)& $e\phi$& $3.3\times 10^{-3}$ & $22.4\times 10^{-3}$& 
(131)(331)& $e\rho$& $2.5\times 10^{-3}$ & $8.6\times 10^{-3}$\\
(131)(332)& $e\ksb$  & $3.5\times 10^{-3}$ & $8.6\times 10^{-5}$& 
(132)(331)& $e\kst$& $2.9\times 10^{-3}$ & $0.126$\\
(132)(332)& $e\phi$& $3.3\times 10^{-3}$ & $0.126$& 
(211)(311)& $\mu\rho$& $4.4\times 10^{-3}$&$6.5\times 10^{-3}$\\
(211)(312)& $\mu\ksb$& $3.6\times 10^{-3}$ & $6.5\times 10^{-3}$& 
(212)(311)& $\mu\kst$& $3.6\times 10^{-3}$ & $6.5\times 10^{-3}$\\
(212)(312)& $\mu\phi$& $3.3\times 10^{-3}$&$6.5\times 10^{-3}$& 
(213)(313)& $\mu\rho$& $4.4\times 10^{-3}$&$6.5\times 10^{-3}$\\
(221)(321)& $\mu\rho$& $4.4\times 10^{-3}$&$93.6\times 10^{-3}$& 
(221)(322)& $\mu\ksb$& $3.6\times 10^{-3}$ & $93.6\times 10^{-3}$\\
(222)(321)& $\mu\kst$& $3.6\times 10^{-3}$ & $0.109$& 
(222)(322)& $\mu\phi$& $3.3\times 10^{-3}$&$0.109$\\
(231)(331)& $\mu\rho$& $4.4\times 10^{-3}$&$81.0\times 10^{-3}$& 
(231)(332)& $\mu\ksb$& $3.6\times 10^{-3}$ & $81.0\times 10^{-3}$\\
(232)(331)& $\mu\kst$& $3.6\times 10^{-3}$ & $0.252$& 
(232)(332)& $\mu\phi$& $3.3\times 10^{-3}$&$0.252$\\
\hline
\end{tabular}
\caption{Bounds on $\l'\l'$ products from $\tau\r \ell + M$ processes.
Only the best bounds are shown.}
\end{center}
\end{table}
%\end{document}

%% file: tab.3
\oddsidemargin=-0.5 true in
%\begin{document}
\def\bd {B_d}
\def\bs {B_s}
\def\t {\times}
\def\r{\rightarrow}

\begin{table}[htbp]
\begin{center}
\begin{tabular}{||c|c|c|c||c|c|c|c||}
\hline 
$\lambda\lambda'$ & Process & Bound & Previous   &
$\lambda\lambda'$ & Process & Bound & Previous   \\
 & & ($\times 10^5$)& bound &
 & & ($\times 10^5$)& bound \\
\hline 
(121)(113)& $\bd\r \mu^\pm e^\mp$ & $2.3$ & $1.0\t 10^{-3}$ &
(121)(123)& $\bs\r \mu^\pm e^\mp$ & $4.7$ & $2.1\t 10^{-3}$\\
(121)(131)& $\bd\r \mu^\pm e^\mp$ & $2.3$ & $9.3\t 10^{-4}$ &
(121)(132)& $\bs\r \mu^\pm e^\mp$ & $4.7$ & $1.4\t 10^{-2}$\\
(121)(213)& $\bd\r e^+ e^-$ & $1.7$ & $2.9\t 10^{-3}$  &
(121)(223)& $\bs\r e^+ e^-$ & $14$ & $1.0\t 10^{-3}$ \\
(121)(231)& $\bd\r e^+ e^-$ & $1.7$ & $8.8\t 10^{-3}$  &
(121)(232)& $\bs\r e^+ e^-$ & $14$ & $7.3\t 10^{-3}$ \\

(122)(113)& $\bd\r \mu^+\mu^-$ & $1.5$ & $5.4\t 10^{-3}$ &
(122)(123)& $\bs\r \mu^+\mu^-$ & $2.7$ & $2.1\t 10^{-3}$\\
(122)(131)& $\bd\r \mu^+\mu^-$ & $1.5$ & $9.3\t 10^{-4}$ &
(122)(132)& $\bs\r \mu^+\mu^-$ & $2.7$ & $1.4\t 10^{-2}$\\
(122)(213)& $\bd\r \mu^\pm e^\mp$ & $2.3$ & $2.9\t 10^{-3}$ &
(122)(223)& $\bs\r \mu^\pm e^\mp$ & $4.7$ & $1.0\t 10^{-2}$\\
(122)(231)& $\bd\r \mu^\pm e^\mp$ & $2.3$ & $8.8\t 10^{-3}$ &
(122)(232)& $\bs\r \mu^\pm e^\mp$ & $4.7$ & $2.7\t 10^{-2}$\\

(123)(113)& $\bd\r \mu^\pm \tau^\mp$ & $62$ & $1.0\t 10^{-3}$ & 
(123)(131)& $\bd\r \mu^\pm \tau^\mp$ & $62$ & $9.3\t 10^{-4}$ \\
(123)(213)& $\bd\r   e^\pm \tau^\mp$ & $49$ & $2.9\t 10^{-3}$ &
(123)(231)& $\bd\r   e^\pm \tau^\mp$ & $49$ & $8.8\t 10^{-3}$\\

(131)(113)& $\bd\r   e^\pm \tau^\mp$ & $49$ & $1.3\t 10^{-3}$ &
(131)(131)& $\bd\r   e^\pm \tau^\mp$ & $49$ & $1.1\t 10^{-3}$\\
(131)(313)& $\bd\r e^+ e^-$ & $1.7$ & $6.8\t 10^{-3}$  & 
(131)(323)& $\bs\r e^+ e^-$ & $14$ & $3.2\t 10^{-3}$  \\
(131)(331)& $\bd\r e^+ e^-$ & $1.7$ & $2.8\t 10^{-2}$  & 
(131)(332)& $\bs\r e^+ e^-$ & $14$ & $2.8\t 10^{-2}$  \\

(132)(113)& $\bd\r \mu^\pm \tau^\mp$ & $62$ & $1.3\t 10^{-3}$ & 
(132)(131)& $\bd\r \mu^\pm \tau^\mp$ & $62$ & $1.2\t 10^{-3}$\\ 
(132)(313)& $\bd\r \mu^\pm e^\mp$ & $2.3$ & $6.8\t 10^{-3}$ & 
(132)(323)& $\bs\r \mu^\pm e^\mp$ & $4.7$ & $4.3\t 10^{-3}$ \\
(132)(331)& $\bd\r \mu^\pm e^\mp$ & $2.3$ & $2.8\t 10^{-2}$ & 
(132)(332)& $\bs\r \mu^\pm e^\mp$ & $4.7$ & $2.8\t 10^{-2}$ \\

(133)(313)& $\bd\r   e^\pm \tau^\mp$ & $49$ & $3.6\t 10^{-5}$  & 
(133)(331)& $\bd\r   e^\pm \tau^\mp$ & $49$ & $2.7\t 10^{-3}$  \\

(231)(213)& $\bd\r   e^\pm \tau^\mp$ & $49$ & $4.1\t 10^{-3}$  & 
(231)(231)& $\bd\r   e^\pm \tau^\mp$ & $49$ & $1.3\t 10^{-3}$  \\
(231)(313)& $\bd\r \mu^\pm e^\mp$ & $2.3$ & $4.2\t 10^{-4}$ & 
(231)(323)& $\bs\r \mu^\pm e^\mp$ & $4.7$ & $3.6\t 10^{-2}$ \\
(231)(331)& $\bd\r \mu^\pm e^\mp$ & $2.3$ & $2.2\t 10^{-2}$ & 
(231)(332)& $\bs\r \mu^\pm e^\mp$ & $4.7$ & $3.2\t 10^{-2}$ \\

(232)(213)& $\bd\r \mu^\pm \tau^\mp$ & $62$ & $4.1\t 10^{-3}$ & 
(232)(231)& $\bd\r \mu^\pm \tau^\mp$ & $62$ & $1.3\t 10^{-2}$ \\
(232)(313)& $\bd\r \mu^+\mu^-$ & $1.5$ & $1.3\t 10^{-2}$ & 
(232)(323)& $\bs\r \mu^+\mu^-$ & $2.7$ & $3.6\t 10^{-2}$ \\
(232)(331)& $\bd\r \mu^+\mu^-$ & $1.5$ & $3.5\t 10^{-3}$ & 
(232)(332)& $\bs\r \mu^+\mu^-$ & $2.7$ & $3.2\t 10^{-2}$ \\

(233)(313)& $\bd\r \mu^\pm \tau^\mp$ & $62$ & $7.7\t 10^{-3}$ & 
(233)(331)& $\bd\r \mu^\pm \tau^\mp$ & $62$ & $3.2\t 10^{-2}$ \\

\hline
\end{tabular}
\caption{$\l\l'$ bounds from leptonic $B_d$ and $B_s$ decays.}
\end{center}
\end{table}
\oddsidemargin=5 pt
%\end{document}

%% file: tab.4
\oddsidemargin=-0.5 true in
%\begin{document}
\def\ksb {\overline{{K^0}^*}}
\def\kst {{K^0}^*}
\def\bd {B_d}
\def\bs {B_s}
\def\t {\times}
\def\r{\rightarrow}

\begin{table}[htbp]
\begin{center}
\begin{tabular}{||c|c|c|c||c|c|c|c||}
\hline
$\lambda'\lambda'$ & Process & Bound & Previous   &
$\lambda'\lambda'$ & Process & Bound & Previous   \\
& & ($\times 10^3$) & bound &
& & ($\times 10^3$) & bound \\
\hline
(111)(213)& $\bd\r \mu^\pm e^\mp$ & $4.7$ & $3\t 10^{-5}$ & 
(111)(313)& $\bd\r   e^\pm \tau^\mp$ & $5.9$ & $5.7\t 10^{-5}$ \\ 
(112)(213)& $\bs\r \mu^\pm e^\mp$ & $9.6$ & $1.2\t 10^{-3}$ & 
(113)(211)& $\bd\r \mu^\pm e^\mp$ & $4.7$ & $1.2\t 10^{-3}$   \\
(113)(212)& $\bs\r \mu^\pm e^\mp$ & $9.6$ & $1.2\t 10^{-3}$   & 
(113)(311)& $\bd\r   e^\pm \tau^\mp$ & $5.9$ & $2.3\t 10^{-3}$  \\ 
(121)(223)& $\bd\r \mu^\pm e^\mp$ & $4.7$ & $9.0\t 10^{-3}$ & 
(121)(323)& $\bd\r   e^\pm \tau^\mp$ & $5.9$ & $22.4\t 10^{-3}$  \\ 
(122)(223)& $\bs\r \mu^\pm e^\mp$ & $9.6$ & $9.0\t 10^{-3}$ & 
(123)(221)& $\bd\r \mu^\pm e^\mp$ & $4.7$ & $7.7\t 10^{-3}$  \\ 
(123)(222)& $\bs\r \mu^\pm e^\mp$ & $9.6$ & $9.0\t 10^{-3}$  &  
(123)(321)& $\bd\r   e^\pm \tau^\mp$ & $5.9$ & $22.4\t 10^{-3}$  \\ 
(131)(233)& $\bd\r \mu^\pm e^\mp$ & $4.7$ & $2.9\t 10^{-3}$ & 
(131)(333)& $\bd\r   e^\pm \tau^\mp$ & $5.9$ & $7.6\t 10^{-3}$\\ 
(132)(233)& $\bs\r \mu^\pm e^\mp$ & $9.6$ & $42\t 10^{-3}$ & 
(133)(231)& $\bd\r \mu^\pm e^\mp$ & $4.7$ & $0.3\t 10^{-3}$  \\ 
(133)(232)& $\bs\r \mu^\pm e^\mp$ & $9.6$ & $0.8\t 10^{-3}$   & 
(133)(331)& $\bd\r   e^\pm \tau^\mp$ & $5.9$ & $0.6\t 10^{-3}$ \\ 
(211)(213)& $\bd\r \mu^+\mu^-$ & $2.1$ & $3.5\t 10^{-3}$  & 
(211)(313)& $\bd\r \mu^\pm \tau^\mp$ & $7.3$ & $6.5\t 10^{-3}$ \\ 
(212)(213)& $\bs\r \mu^+\mu^-$ & $3.9$ & $3.5\t 10^{-3}$ &
(213)(311)& $\bd\r \mu^\pm \tau^\mp$ & $7.3$ & $6.5\t 10^{-3}$ \\ 
(221)(223)& $\bd\r \mu^+\mu^-$ & $2.1$ & $1.4\t 10^{-3}$  & 
(221)(323)& $\bd\r \mu^\pm \tau^\mp$ & $7.3$ & $93.6\t 10^{-3}$ \\ 
(222)(223)& $\bs\r \mu^+\mu^-$ & $3.9$ & $2.7\t 10^{-3}$ &
(223)(321)& $\bd\r \mu^\pm \tau^\mp$ & $7.3$ & $0.109$  \\ 
(231)(233)& $\bd\r \mu^+\mu^-$ & $2.1$ & $27\t 10^{-3}$ & 
(231)(333)& $\bd\r \mu^\pm \tau^\mp$ & $7.3$ & $81\t 10^{-3}$ \\  
(232)(233)& $\bs\r \mu^+\mu^-$ & $3.9$ & $84\t 10^{-3}$ &
(233)(331)& $\bd\r \mu^\pm \tau^\mp$ & $7.3$ & $67.5\t 10^{-3}$ \\  
\hline
\end{tabular}
\caption{$\l'\l'$ bounds from leptonic $B_d$ and $B_s$ decays.}
\end{center}
\end{table}
\oddsidemargin=5 pt
%\end{document}

%% file: tab.5
%\documentstyle[11pt]{article}
%\pagestyle{empty}
%\oddsidemargin=-0.5 true in
%\begin{document}
\def\ksb {\overline{{K^0}^*}}
\def\kst {{K^0}^*}
\def\bd {B_d}
\def\bs {B_s}
\def\t {\times}
\def\r{\rightarrow}

\begin{table}[htbp]
\begin{center}
\begin{tabular}{||c|c|c||c|c|c||}
\hline
$\lambda'\lambda'$ & Final & Bound & $\lambda'\lambda'$ & Final & Bound \\
& state & & & state & \\
\hline
(111)(213)& $\pi \mu^\pm e^\mp$ & $5.0\t 10^{-4}$ & 
(113)(211)& $\pi \mu^\pm e^\mp$ & $5.0\t 10^{-4}$ \\
(121)(223)& $\pi \mu^\pm e^\mp$ & $5.0\t 10^{-4}$ & 
(123)(221)& $\pi \mu^\pm e^\mp$ & $5.0\t 10^{-4}$ \\
(131)(233)& $\pi \mu^\pm e^\mp$ & $5.0\t 10^{-4}$ & 
(133)(231)& $\pi \mu^\pm e^\mp$ & $5.0\t 10^{-4}$ \\
(112)(213)& $  K \mu^\pm e^\mp$ & $3.3\t 10^{-4}$ & 
(113)(212)& $  K \mu^\pm e^\mp$ & $3.3\t 10^{-4}$ \\
(122)(223)& $  K \mu^\pm e^\mp$ & $3.3\t 10^{-4}$ & 
(123)(222)& $  K \mu^\pm e^\mp$ & $3.3\t 10^{-4}$ \\
(132)(233)& $  K \mu^\pm e^\mp$ & $3.3\t 10^{-4}$ & 
(133)(232)& $  K \mu^\pm e^\mp$ & $3.3\t 10^{-4}$ \\
(212)(213)& $  K \mu^+\mu^-$    & $9.7\t 10^{-5}$ &
(222)(223)& $  K \mu^+\mu^-$    & $9.7\t 10^{-5}$ \\
(232)(233)& $  K \mu^+\mu^-$    & $9.7\t 10^{-5}$ &
(112)(113)& $  K   e^+  e^-$    & $9.6\t 10^{-5}$ \\
(122)(123)& $  K   e^+  e^-$    & $9.6\t 10^{-5}$ &
(132)(133)& $  K   e^+  e^-$    & $9.6\t 10^{-5}$ \\
\hline
\end{tabular}
\caption{$\l'\l'$ bounds from semileptonic $B_d$ decays.}
\end{center}
\end{table}
%\end{document}

%% file: tab.6
%\documentstyle[11pt]{article}
%\pagestyle{empty}
%\oddsidemargin=-0.5 true in
%\begin{document}
\def\ksb {\overline{{K^0}^*}}
\def\kst {{K^0}^*}
\def\bd {B_d}
\def\bs {B_s}
\def\t {\times}
\def\r{\rightarrow}

\begin{table}[htbp]
\begin{center}
\begin{tabular}{||c||c||c||c||}
\hline
& Ali {\em et al} \protect\cite{ali} & Greub {\em et al} 
\protect\cite{greub} & Melikhov {\em et al} \protect\cite{melikhov} \\
\hline
     & CC: $2.4\t 10^{-5}$   & CC: $9.5\t 10^{-5}$ &   CC: $6.4\t 10^{-5}$\\
BaBar & CD: $1.7\t 10^{-5}$  & CD: Not possible    &   CD: Not possible\\
     & Inc: $9.7\t 10^{-5}$  & Inc: $2.1\t 10^{-4}$ & Inc: $1.7 \t 10^{-4}$\\
\hline
     & CC: $1.3\t 10^{-4}$   & CC: $2.4\t 10^{-4}$ &   CC: $2.0\t 10^{-4}$\\
Belle & CD: $1.2\t 10^{-5}$  & CD: Not possible    &   CD: Not possible\\
     & Inc: $2.6\t 10^{-4}$  & Inc: $3.9\t 10^{-4}$ & Inc: $3.3 \t 10^{-4}$\\
\hline
\end{tabular}
\caption{Bounds on $|\l'_{2i2}\l'_{2i3}|$ from the decay $B\r K\mu^+\mu^-$ 
with different schemes for evaluating the SM branching ratio and two sets
of experimental values from BaBar and Belle. CC and CD stand for complete
constructive and destructive interference between SM and RPV amplitudes
respectively, and Inc means incoherent amplitude sum. For two theoretical
schemes, a complete destructive interference is ruled out from the data.}
\end{center}
\end{table}
%\end{document}